Review Article

# Genes predisposing to type 1 diabetes mellitus and pathophysiology: a narrative review

Tajudeen Yahaya,[1] Titilola Salisu[2]

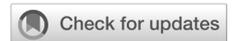

**ABSTRACT**

The possibility of targeting the causal genes along with the mechanisms of pathogenically complex diseases has led to numerous studies on the genetic etiology of some diseases. In particular, studies have added more genes to the list of type 1 diabetes mellitus (T1DM) suspect genes, necessitating an update for the interest of all stakeholders. Therefore this review articulates T1DM suspect genes and their pathophysiology. Notable electronic databases, including Medline, Scopus, PubMed, and Google-Scholar were searched for relevant information. The search identified over 73 genes suspected in the pathogenesis of T1DM, with human leukocyte antigen, insulin gene, and cytotoxic T lymphocyte-associated antigen 4 accounting for most of the cases. Mutations in these genes, along with environmental factors, may produce a defective immune response in the pancreas, resulting in β-cell autoimmunity, insulin deficiency, and hyperglycemia. The mechanisms leading to these cellular reactions are gene-specific and, if targeted in diabetic individuals, may lead to improved treatment. Medical practitioners are advised to formulate treatment procedures that target these genes in patients with T1DM.

**KEYWORDS** autoimmunity, hyperglycemia, insulin, pancreas, type 1 diabetes mellitus

Diabetes mellitus (DM) begins when hyperglycemia occurs. A hormone synthesized by the pancreatic β-cells, known as insulin, regulates blood glucose by transporting it from the bloodstream into the cells where it is metabolized. Another hormone produced by the pancreatic α-cells, known as glucagon, also assists the body in the blood glucose regulation process. In most cases, type 1 diabetes mellitus (T1DM) occurs when the immune system, which normally attacks only foreign antigens, faultly destroys the β-cells. Consequently, the pancreas stops making insulin, leading to the retention of glucose in the blood.[1] Some people develop a condition known as secondary DM, which is like T1DM, but the immune system does not destroy the β-cells.[2] Instead, the cells are removed by certain stimuli in the pancreas, as indicated by histopathological changes.[2] T1DM was previously named juvenile DM but was renamed because adults can also develop the disease. Moreover, T1DM is not the only form of DM that affects children and young adults. Particularly, type 2 diabetes mellitus (T2DM) is becoming increasingly prevalent among young individuals as more and more children are becoming obese and overweight nowadays.³ Maturity-onset diabetes of the young and neonatal diabetes mellitus are other forms of DM that mainly affect children and neonates with different clinical features from T1DM.

T1DM is the most frequent chronic illness among children, accounting for 5–10% of all DM cases.² It is the most burdensome form of DM, with a decreased life expectancy.⁴ About 30 million people live with T1DM worldwide, and the prevalence is predicted to increase three-fold in the 2040s.⁵ Uncontrolled T1DM may cause chronic complications, including retinopathy, nephropathy, neuropathy, and vascular diseases. T1DM has a strong genetic etiology modulated by some environmental triggers that may prevent the disease if avoided before autoimmunity. The treatment involves constant insulin injections, which is expensive; thus, a better approach is necessary to stem the burden of the disease. As autoimmunity is the hallmark of T1DM, an intervention that re-programs the immune system may be a potential way to treat the disease. This observation suggests that precision medicine that targets causal genes and the pathophysiology of complicated diseases, such as DM, may provide a future management when fully understood. Hence, this review was initiated to search notable electronic databases, including Medline, Scopus, PubMed, and Google-Scholar to establish T1DM suspected genes. The result will facilitate the formulation of drugs and treatment procedures that target these genes and their mechanisms.

## T1DM SUSPECT GENES

At least 18 regions in the human genome were identified as hotspots for T1DM using genome-wide linkage analysis. These regions host many genes and are labeled *IDDM1-IDDM18* because T1DM was formerly called insulin-dependent DM (IDDM).⁶ However, improvement in biological techniques has led to a greater understanding of the genetic basis of T1DM. More precise studies have reaffirmed the roles of some previously suspected genes in T1DM pathogenesis and new ones have been discovered. Many genes have been reportedly linked with T1DM, but only those with clear mechanisms are presented in this review. Tables 1–3 show the update on T1DM candidate genes, grouped according to their mechanisms of action, which are disruption of T cell regulatory activities, increased susceptibility to microbial infection, oxidative stress, cytokine-induced β-cell loss, and β-cell reconfiguration. However, more genes are being added continually.

**Most frequently suspected T1DM predisposing genes**

Of the genes described in Tables 1–3, mutations in the major histocompatibility complex (MHC), insulin gene (*INS*), cytotoxic T lymphocyte-associated protein 4 (*CTLA4*), protein tyrosine phosphatase, non-receptor type 2 (*PTPN2*), and protein tyrosine phosphatase, non-receptor type 22 (*PTPN22*) genes account for most of the T1DM cases. Detailed information on the diabetogenic activities of these genes is given below.

**Table 1.** T1DM suspect genes that disrupt T cell immune regulatory activities and increase susceptibility to microbial infection

| Gene | Full name | Locus | Mechanism |
|---|---|---|---|
| CTLA4 | Cytotoxic T-lymphocyte associated protein 4 | 2q33.2 | Causes massive lymphocyte proliferation, resulting in overexpression of T cells, which attack self-antigens, leading to β-cell apoptosis.⁷ |
| FOXP3 | Forkhead box P3 | Xp11.23 | Disrupts the activities of many endocrine glands, affecting the immune regulatory T cells and setting of autoimmunity.⁸ |
| PRKCQ | Protein kinase C theta | 10p15.1 | Loss of T cell signaling, leading to a compromised immune system.⁹ |
| IL7R | Interleukin 7 receptor | 5p13.2 | Causes a compromised immune system, characterized by depletion of T cells.¹⁰ |
| IL2 | Interleukin 2 | 4q27 | Causes functional loss of T cell (Treg) regulation, resulting in autoimmunity.¹¹ |
| ITGB7 | Integrin subunit beta 7 | 12q13.13 | It disrupts the activities of T cells in the pancreas, resulting in autoimmunity.¹² |
| SH2B3 | SH2B adaptor protein 3 | 12q24.13 | Increases self-reactive T lymphocyte proliferation.¹³ |
| RASGRP1 | RAS guanyl releasing protein 1 | 15q14 | Reduced expression results in splenomegaly and autoantibodies, while complete deletion decreases T cell development in the thymus.¹⁴ |

**Table 1.** (continued)

| Gene | Full name | Locus | Mechanism |
| --- | --- | --- | --- |
| *CD28* | CD28 molecule | 2q33.2 | Mediates loss of function of regulatory T cells, leading to autoimmunity of the pancreatic β-cells.[15] |
| *ICOS* | Inducible T cell costimulator | 2q33.2 | It suppresses insulin-specific regulatory T cells.[16] |
| *ART2/RT6* | Antisense to ribosomal RNA transcript protein 2 | 11p15.4 | Causes insulitis in which the islets are massively infiltrated by macrophages and T cells.[17] |
| *UBASH3A* | Ubiquitin associated and SH3 domain containing A | 21q22.3 | Increases UBASH3A expression in CD4+ T cells, disrupting NF-κB signaling, which in turn represses *IL2* gene expression.[18] |
| *SUMO4* | Small ubiquitin-like modifier 4 | 6q25.1 | Reduces sumoylation capacity, causing overexpression of NF-κB and IL12B, and leading to β-cell autoimmunity.[19] |
| MHC | Major histocompatibility complex | 6p21.31 | Causes loss of T cell signaling, leading to autoimmunity of the pancreatic β-cells.[20] |
| NRP1 | Neuropilin 1 | 10p11.22 | Causes defects in Treg phenotype decreasing its signaling activities, culminating in autoimmunity of the β-cells.[21] |
| MIF | Macrophage migration inhibitory factor | 22q11.23 | Causes high MIF concentrations in the blood, resulting in dysfunctional pancreatic islets. Also causes loss of function of immune cells, such as macrophages and T cells, increasing the risk of β-cell autoimmunity.[22] |
| *CD226* | Cluster of differentiation 226 | 18q22.2 | Increases the frequency of GAD autoantibody and lowers C-peptide levels, accelerating β-cell destruction.[23] |
| *IL6* | Interleukin 6 | 7p15.3 | Impairs immune response, leading to β-cell autoimmunity, it also leads to impaired fat breakdown in tissues such as muscle cells.[24] |
| *IL10* | Interleukin 10 | 1q32.1 | It negatively regulates a T cell called Th17, setting up autoimmunity of the pancreatic β-cells.[25] |
| *IL12B* | Interleukin 12B | 5q31-33 | It negatively regulates a T cell called Th17, setting up autoimmunity of the pancreatic β-cells.[26] |
| *PTPN22* | Protein tyrosine phosphatase, non-receptor type 22 | 1p13.2 | Disrupts elimination of autoreactive β-cells, and upregulates certain β-cell genes, leading to β-cell death.[27] |
| *VDR* | Vitamin D receptor | 12q13.11 | It causes dysfunctional immune mechanism, leading to autoimmunity.[28] |
| *IL26* | Interleukin 26 | 12q15 | Disrupts sensing of bacteria by the defense mechanism,[29] resulting in the attack of β-cells. |
| *TAGAP* | T cell activation RhoGTPase activating protein | 6q25.3 | Causes deficiency of several important regulators of antiviral response, including IFN-β, culminating in loss of antiviral signaling pathways, and β-cell autoimmunity.[30] |
| *TYK2* | Tyrosine kinase 2 | 19p13.2 | Represses TYK2, predisposing the body to virus attack.[31] |
| *GAD2* | Glutamic acid decarboxylase 2 | 10p12.1 | Promotes invasion of β-cells by viruses, setting of autoimmunity of the β-cells, leading to insulin deficiency.[32] |
| *OAS1* | 2'-5'-oligoadenylate synthetase 1 | 12q24.13 | Predisposes to viral infection, setting in β-cell autoimmunity.[33] |
| *IFIH1* | Interferon induced with helicase C domain 1 | 2q24.2 | Produces abnormal antiviral immune response, increasing the risk of autoimmunity.[34] |
| *FUT2* | Fucosyltransferase 2 | 19q13.33 | Alters the gut microbiome, increasing the risk of autoimmunity of the β-cells.[35] |

T1DM=type 1 diabetes mellitus; Treg=regulatory T cell; NF-κB=nuclear factor kappa B; Th17=T helper 17 cells; IFN-β=interferon beta

**Table 2.** T1DM suspect genes that mediate oxidative stress and cytokine-induced β-cell loss

| Gene | Full name | Locus | Mechanism |
|---|---|---|---|
| SIRT1 | Sirtuin 1 | 10q21.3 | Produces defective sirtuin, inducing cytokines to attack wrong cells, particularly β-cells.[36] |
| ERBB3 | Erb-B2 receptor tyrosine kinase 3 | 12q13.2 | Disrupts antigen presentation, causing cytokine-induced β-cell apoptosis.[37] |
| PTPN2 | Protein tyrosine phosphatase, non-receptor type 2 | 18p11.21 | Produces cytokine, which induces destruction of β-cells and mitochondrial apoptotic pathway, disrupting glucose metabolism.[38] |
| BACH2 | BTB domain and CNC homolog 2 | 6q15 | Decreases *PTPN2* expression, promoting pro-inflammatory cytokine-induced β-cell apoptosis.[39] |
| IL21 | Interleukin 21 | 4q27 | Causes inflammatory infiltration of pancreatic islets.[40] |
| CENPW/C6orf173 | Centromere protein W | 6q22.32 | Contributes to the generation of autoantibodies in the islets.[41] |
| AFF3 | AF4/FMR2 family member 3 | 2q11.2 | Alters the expression of monocytes as well as macrophages and dendritic cells, causing gradual destruction of β-cells.[42] |
| TNFAIP3 | Tumor necrosis factor alpha induced protein 3 | 6q23.3 | Loss of functional regulation of β-cell apoptosis, leading to poor glycemic control.[43] |
| RNLS | Renalase, FAD dependent amine oxidase | 10q23.31 | Alters the expression of monocytes as well as macrophages and dendritic cells, causing gradual destruction of β-cells.[44] |
| GPR183/EBI2 | G protein-coupled receptor 183 | 13q32.3 | Changes the expression of monocytes as well as macrophages and dendritic cells, causing gradual destruction of β-cells.[42] |
| CLECL1 | C-type lectin like 1 | 12p13.31 | Alters the expression of monocytes as well as macrophages and dendritic cells, causing gradual destruction of β-cells.[42] |
| DEXI | Dexamethasone-induced protein | 16p13.13 | Depletes gut microbial metabolites, causing autoimmunity of β-cells.[45] |
| SUOX | Sulfite oxidase | 12q13.2 | Alters the expression of monocytes as well as macrophages and dendritic cells, causing gradual destruction of β-cells.[42] |
| SMARCE1 | SWI/SNF related, matrix associated, actin dependent regulator of chromatin, subfamily e, member 1 | 17q21.2 | Alters the expression of monocytes as well as macrophages and dendritic cells, causing gradual destruction of β-cells.[42] |
| FKRP | Fukutin-related protein | 19q13.32 | Alters the expression of monocytes as well as macrophages and dendritic cells, causing gradual destruction of β-cells.[42] |
| IL27 | Interleukin 27 | 16p11.2 | Promotes inflammations of the islet cells.[46] |
| SOD2 | Superoxide dismutase 2 | 6q25.3 | Enhances generation of free oxygen radical, stressing, and contributing to β-cell death.[6] |
| BCL-2 | B-cell leukemia/lymphoma 2 | 18q21.33 | Initiates intracellular apoptotic pathway, resulting in islet cell death.[47] |
| FAS/CD95 | Fas cell surface death receptor | 10q23.31 | Initiates glucose-induced DNA fragmentation, resulting in pancreatic islet apoptosis.[48] |
| IL-1A | Interleukin 1 alpha | 2q14.1 | Initiates glucose-induced DNA fragmentation, resulting in pancreatic islet apoptosis.[48] |
| APOC3 | Apolipoprotein C3 | 11q23.3 | Causes influx of $Ca^{2+}$ into the β-cells, destroying the cells.[49] |
| BAD | BCL2-associated agonist of cell death | 11q13.1 | Loss of control of programmed cell death.[50] |
| NOS2 | Nitric oxide synthase 2 | 17q11.2 | Produces nitric oxide, inducing apoptosis in several cells, including β-cells.[51] |

**Table 2.** (continued)

| Gene | Full name | Locus | Mechanism |
| --- | --- | --- | --- |
| PTEN | Phosphatase and tensin homolog | 10q23.31 | Causes gradual destruction of β-cells.[52] |
| CASP7 | Caspase 7 | 10q25.3 | Causes gradual destruction of β-cells.[53] |
| HIP14 | Huntingtin-interacting protein 14 | | Accelerates apoptotic cell death.[54] |
| STAT4 | Signal transducer and activator of transcription 4 | 2q32.2 | Its activation causes cytokine-induced islet apoptosis, resulting in β-cell dysfunction.[55] |
| CCR5 | C-C chemokine receptor 5 | 3p21.31 | Mediates infiltration of the pancreatic islets by abnormal blood cells, culminating in the gradual destruction of β-cells.[56] |
| ITPR3 | Inositol 1,4,5-trisphosphate receptor type 3 | 6p21.31 | Causes dysregulated calcium transport, leading to loss of insulin secretion function of the β-cells, culminating in high blood glucose and high HbA1c beyond healthy levels.[57] |

T1DM=type 1 diabetes mellitus; HbA1c=hemoglobin A1c

**Table 3.** T1DM suspect genes that reconfigure β-cell morphology and identity

| Gene | Full name | Locus | Mechanism |
| --- | --- | --- | --- |
| INS | Insulin gene | 11p15.5 | Breakage of phenylalanine amino acid, hampering maturation of preproinsulin to insulin.[58] |
| HNFIA | Hepatocyte nuclear factor-1 alpha | 12q2431 | Disruption of insulin biosynthesis, leading to hyperglycemia and osmotic diuresis.[59] |
| SLC2A2 | Solute carrier family 2 member 2 | 3q26.2 | Fanconi-Bickel syndrome, causing lack of insulin in utero.[60] |
| Amylin | Islet amyloid polypeptide | 12p12.1 | Causes over secretion of amylin, which aggregate in the pancreatic β-cells, causing β-cell death and impairing insulin secretion.[61] |
| NKX2.2 | NK2 homeobox 2 | 20p11.22 | Causes β-cell reconfiguration, leading to loss of insulin synthesis.[62] |
| GLIS3 | GLI-similar 3 | 9p24.2 | Suppresses insulin gene expression, causing pancreatic β-cell death.[63] |
| NEUROD1 | Neuronal differentiation 1 | 2q32 | Affects β-cell morphogenesis and differentiation, promoting β-cell apoptosis.[64] |
| CTSH | Cathepsin H | 15q25.1 | Progressively reduces the survival of β-cells.[65] |
| DLL1 | Delta-like1-Drosophila | 6q27 | Hampers β-cell differentiation and maturation.[66] |
| ORMDL3 GSDMB | Orosomucoid like 3 | 17q12 | Reduces sulfatide levels in the pancreatic islet, causing unregulated proinsulin folding as well as apoptosis.[67] |
| FADD | Fas-associated death domain | 11q13.3 | Disrupts transcription of pancreatic cells.[6] |
| CDKN1B/p27KIP1 | Cyclin-dependent kinase inhibitor 1B | 12p13.1 | Causes loss of β-cell mass, and poor glucose metabolism.[68] |
| IRS2 | Insulin receptor substrate 2 | 13q34 | Reduces β-cell mass.[69] |
| IL2RA/CD25 | Interleukin 2 receptor alpha | 10p15.1 | Causes inflammatory bowel disease, β-cell enlargement, leading to autoimmunity.[70] |

T1DM=type 1 diabetes mellitus

**MHC**

MHC is about a 4 megabases genetic section on chromosome 6 (6p21) containing immune recognition genes.[20] Many versions of MHC occur among animal species.[20] In humans, it is called the HLA complex and contains over 200 genes grouped into class I, II, and III.[20] Class I has three major genes, named HLA-A, HLA-B, and HLA-C.[71] The proteins encoded by these genes are expressed on the surface of most nucleated cells, where they bind to protein fragments (peptides) exported from within the cell.[71] Class I proteins present these peptides to the immune system, and apoptosis is induced if the peptides are recognized as foreign, such as viral or bacterial peptides.[71] Human MHC class II has six main genes, including HLA-DPA1, HLA-DPB1, HLA-DQA1, HLA-DQB1, HLA-DRA, and HLA-DRB1.[71] The proteins produced by these genes are expressed mainly on the surface of certain immune system cells, including B lymphocytes, dendritic cells, macrophages, and activated T lymphocytes.[71] Class II proteins display peptides to the immune system.[71]

HLA class I and II genes are highly polymorphic, and inheritance of certain variants may predispose the affected individual to an autoimmune disorder. Certain HLA class II alleles or combinations of alleles (haplotypes) significantly increase the risk of T1DM, while others confer reduced or protective effects.[72] For example, most individuals with T1DM are carriers of either HLA-DR3, DQB1*0201 (also called DR3-DQ2), or DR4-DQB1*0302 (otherwise known as DR4-DQ8).[72] Also, inheriting the HLA haplotype DRB1*0302-DQA1*0301, particularly when combined with DRB1*0201-DQA1*0501, increases genetic predisposition to T1DM as much as 20-fold.[72] In contrast, the haplotype DRB1*0602-DQA1*0102 rarely predisposes to T1DM,[72] while the haplotype HLA-DQ6 (HLA-DQA1*0102-DQB1*0602) is protective.[6]

**INS gene**

INS gene was the second gene linked with T1DM and accounts for about 10% of T1DM cases.[34] It encodes the precursor to insulin, which assists the body to store energy for later use.[73] For example, insulin helps the body store glucose as glycogen or fat rather than metabolizing it. Insulin has two separate polypeptide chains, chains A and B, and are connected by disulfide bonds.[73] Unlike some proteins that are synthesized by several genes, insulin is synthesized by the INS gene only. Some animals, such as rats and mice, have two insulin genes but humans only have one. To produce insulin, the INS gene secretes an inactive insulin precursor called preproinsulin, which is converted to another inactive substance called proinsulin by removal of a signaling peptide.[73] Insulin is finally produced from proinsulin by removal of the C-peptide that binds chains A and B together.[73] However, mutations in the INS gene may disrupt the insulin biosynthetic network, causing some diseases. A point mutation in the INS gene known as C96Y synthesizes mutant proinsulin, resulting in endoplasmic reticulum (ER) stress, which progressively causes death of β-cells and T1DM.[74] Several other point mutations have also been reported in individuals and animals with T1DM.

The promoter region of the INS gene is highly polymorphic and contains several variable numbers of tandem repeats (VNTRs), grouped into classes I, II, and III.[20] VNTR I contains 26–63 repeats, VNTR II has 80 repeats, and VNTR III has 140–210 repeats. Among Caucasians, VNTR I is highly prevalent, VNTR III is moderately prevalent, while VNTR II is rare. VNTR I homozygotes often develop T1DM than VNTR III, while VNTR II confers a protective effect.[20] A polymorphism in the promoter region of the INS gene determines the regulatory activities of the transcription factor autoimmune regulator on thymic expression of insulin. The VNTR I allele reduces tolerance to insulin and its precursors, repressing insulin transcription and predisposing the individual to T1DM.[20] The VNTR II allele promotes the expression of insulin mRNA in the thymus.[20]

**CTLA4 gene**

CTLA4 gene is expressed on the surface of activated T cells, where it attenuates the immune response by binding to ligands CD80 or CD86 expressed on the surface of antigen-presenting cells.[75] The CTLA4-CD80/CD86 complex represses the IL-2 receptor (CD25), reducing IL-2 synthesis or triggering cell death in already activated cells.[75] CTLA4 protein also mediates the suppressive activity of CD4+ CD25+ T regulatory cells.[75] The expression of CTLA4 in activated T lymphocytes shows that it maintains immune function by preventing the inflammatory response and autoimmunity.[74] Thus, CTLA4 plays a negative regulatory role in immune function by preventing the overexpression of T cells. Functional impairment in the CTLA4 gene may cause overexpression of T cells, causing it to attack self-antigens. Deletion of the

*CTLA4* gene in mice leads to massive proliferation of lymphocytes, resulting in autoimmunity and death.[74]

Several autoimmune disorders have been associated with functional loss or impairment of the *CTLA4* gene. A single nucleotide polymorphism (SNP) +6230G>A characterized by splicing of the gene or altered mRNA is linked with an increased risk for T1DM.[75] The promoter region of the *CTLA4* gene is polymorphic and a particular SNP −319C>T, which reduces the transcription of the gene, is related to a high risk of T1DM.[75] SNP involving an A-to-G substitution at nucleotide 49 in exon 1 (G49A), causing an amino acid substitution (Thr17Ala) has also been reported in patients with T1DM.[75] The predisposing Ala17 allele is partially glycosylated in the ER, resulting in retrograde transport of some molecules into the cytoplasm for lysis,[75] leading to less *CTLA4* (Ala17) at the cell surface, which may be responsible for loss of function of the *CTLA4* gene expressed by individuals with the +49G allele.[75] Thus, the G49A allele reduces the negative regulatory role of *CTLA4* and predisposes to TIDM, compared with the 49 G/G alleles, which confers protection.[74]

### *PTPN22* and *PTPN2* genes

*PTPN22* and *PTPN2* genes both code for protein tyrosine phosphatase (PTP) signaling molecules that modulate and regulate several biological processes, including cell growth, survival, and differentiation.[76] PTPs are important regulators of signaling transduction, as they relay signals from the cell into the nucleus.[76] Besides cell growth and differentiation, these molecules initiate cell signaling for T cell immune regulatory activities.[76] PTPs are so important in immune regulation that they are more expressed on immune cells than other body tissues.[76] A deficiency of PTP in mice distinctively upregulates immune status with severe abnormalities in hematopoiesis, suggesting that PTPs play an important role in maintaining a balanced immune system.[76] A deficiency in these signaling molecules because of inactivation or loss of *PTPN2* or *PTPN22* leads to decreased suppression of the inflammatory response resulting from reduced negative regulation.[76]

*PTPN22* is the fourth gene linked with T1DM in which the rs2476601 SNP disrupts PTP intracellular signaling, leading to loss of negative immune regulation.[75] The SNP causes a single substitution of arginine for tryptophan in the encoded protein (R620W), leading to a decrease in T cell and B cell receptor signaling.[77] This may disrupt tolerance in both T and B cells, ultimately resulting in diabetes-specific autoimmunity.[77] The autoimmunity induced by this SNP is characterized by a preponderance of autoreactive B cells and autoantibodies, both of which are biomarkers for the onset and progression of T1DM.[77] The mechanism for the role of the *PTPN2* gene in the pathogenesis of T1DM is complex, consequent to expression in many cells, but is suspected to involve destruction of pancreatic β-cells. Functional loss of the *PTPN2* gene may also hamper negative regulation of the apoptotic pathway, leading to overexpression of T cells. Repression of the *PTPN2* gene impairs insulin production by β-cells in diabetic mice. In normal individuals, *PTPN2* gene blocks insulin signaling by dephosphorylation of its β-chain receptor with the assistance of PTP1B phosphatase. This, in turn, controls gluconeogenesis in the liver by suppressing STAT3 signaling and decreasing glucose production. Mutant *PTPN2* induces mitochondrial apoptotic pathways, resulting in β-cell apoptosis and unbalanced glucose metabolism. A version of *PTPN2* (rs1893217) has been reported to upregulate T cell receptor signaling in mice, causing impaired self-antigen recognition and β-cells destruction consequent to the loss of negative regulation.[38]

### Application of autoantibodies and proteins of T1DM suspect genes as biomarkers

No guidelines have been established for T1DM genetic testing as it is done in DM with monogenetic etiology. This is partly because several genes may interact to cause T1DM and testing for individual genes may not be cost-effective. Also, genetic susceptibility alone may not fully explain the etiology of the disease as environmental triggers, such as diet, infection, and pollutants may play a role in onset of the disease. Nevertheless, the Immunology of Diabetes Society has suggested that certain autoantibodies in individuals with a family history of T1DN may predict the likelihood of the disease. These autoantibodies include islet cell antibodies, insulin autoantibodies, the GAD autoantibody, and the protein tyrosine phosphatase IA-2/ICA512.[78] Tables 1 and 2 present some genes that produce these autoantibodies. The β-cell function determined by the first-phase insulin response in the intravenous glucose tolerance test can also predict onset of the disease.[78] The genes whose mutations or variants affect β-cell differentiation and morphology, resulting in loss or reduced functions, are shown in Table 3. The levels of the proteins produced by genes predisposing to enteroviruses, such as B coxsackieviruses, are also a good biomarker of the disease. Notable among these genes are the HLA, *PPTN22*, *PPTN2*, and *CTLA4* genes discussed earlier. Abnormal levels of the MHC and CTLA4 proteins could indicate enteroviral induced autoimmunity in T1DM, prompting a therapeutic measure. Viral infection may also raise the concentration of interferons in islet β-cells, which overexpress MHC I, resulting in increased susceptibility to cytotoxic CD8+ T cell recognition and destruction.[79] Other genes listed earlier whose concentrations are sensitive to viral infections include *ERBB3*, *PRKCQ*, *IL7R*, *IL2*, *TAGAP*, *IL26*, *IL10*, *IL12B*, and *OAS1* (Table 1).

In summary, the reviewed articles indicate that mutations in or variants of certain genes may induce autoimmunity in β-cells by compromising the immune system, predisposing the carrier to T1DM. Many T1DM predisposing genes have been identified; however, individual genes alone cannot cause the disease. These genes interact with each other, suppressing or overexpressing the functions of one another and may induce DM along with environmental stimuli. These genes initiate DM through different mechanisms, which may result in improved treatment. Healthcare providers are advised to formulate treatment plans that target these genes and mechanisms rather than the current generalized treatment procedure.

**Conflict of Interest**
The authors affirm no conflict of interest in this study.

**Acknowledgment**
None.

**Funding Sources**
None.